\documentclass[11pt,a4paper]{article}
\usepackage{graphicx}
\begin{document}

\Large 
\centerline{\bf Comment on Japanese Detection of Air Fluorescence Light} 
\centerline{\bf from a Cosmic Ray Shower in 1969}
\vspace{5mm}
\large
\centerline{Bruce R. Dawson}
\centerline {School of Chemistry \& Physics, University of Adelaide, Adelaide 5005 Australia}
\normalsize

\begin{abstract}
We examine the claim made by Hara et al.\cite{Hara} in 1969 of the
observation of a $10^{19}$eV cosmic ray extensive air shower using the
air fluorescence technique.  We find that it is likely that
fluorescence light was observed, confirming this as the first such
observation.  The work of Hara et al. and their friendly competitors
at Cornell University paved the way for modern experiments like the
Pierre Auger Observatory and the Telescope Array.
\end{abstract}

\section{Introduction}
Investigations into the feasibility of detecting air fluorescence
light from extensive air showers were conducted in the 1960's by
groups led by Suga in Japan and Greisen in the United States.  Results
from the Japanese experiment, reported by Hara et al.\cite{Hara} in
1969, are reviewed here.  In that report, the authors say ``One event
is very likely due to the atmospheric scintillation [fluorescence]
light from an air shower whose primary energy and distance are about
$10^{19}$eV and 3\,km, respectively''.  This was the first reported
observation of fluorescence light from an air shower.  The purpose of
the present short note is to the review this observation in the light
of our modern understanding of fluorescence detection.

The Japanese experiment ran at the Dodaira Observatory (altitude
876\,m) for a period of 5 months from December 1968.  The fluorescence
telescope consisted of a 1.6\,m diameter Fresnel lens focussing light
onto a camera of 24 PMTs, each of which imaged a 4.5$^\circ$ degree
portion of the sky.  For the observation described here, the telescope
with its field of view of $23^\circ \times 20^\circ$ was centered at
an elevation of 30$^\circ$.  The design was similar to Greisen's
Cornell telescope \cite{Cornell}.  However the Japanese design had the
advantage of a larger Fresnel lens, and faster electronics.  The rise
and fall-times of pulses on the cathode-ray tube displays were
0.12$\mu$s and 0.2$\mu$s respectively.

The potential fluorescence observation (event \#12 in Fig 3 of
\cite{Hara}) triggered 8 PMTs with an angular track length of
$18.4^\circ$ and a duration of 1.9$\mu$s.  In the next section we
review the event geometry before considering the shape of the light
profile received at the telescope.

\section{Event Geometry}

\begin{figure}[t]
\begin{centering}
\includegraphics[width=10cm]{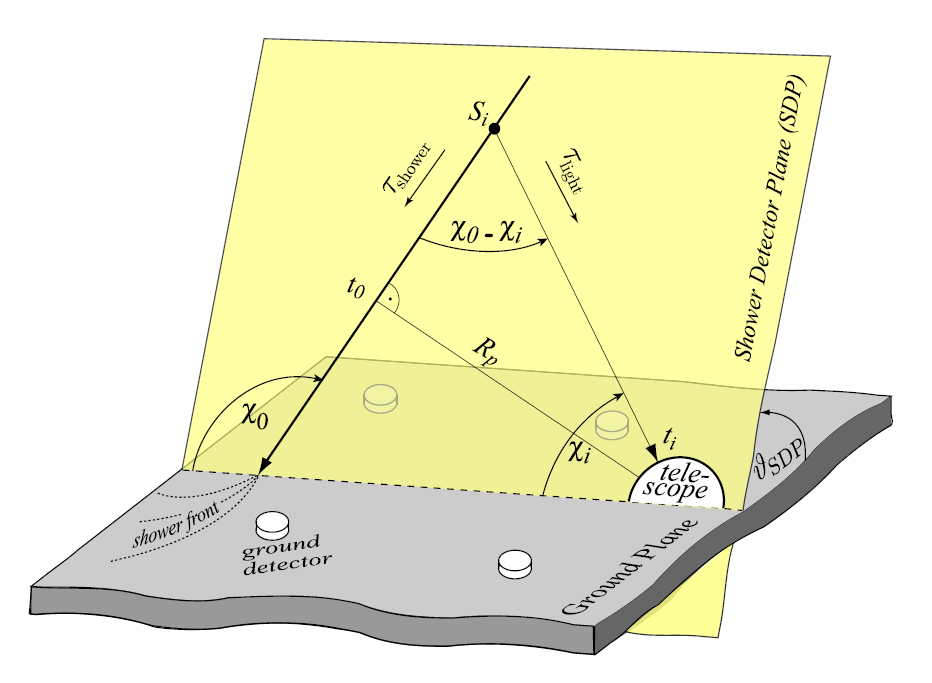}
\caption{The shower axis and the telescope define the shower-detector
  plane (SDP).  The timing fit returns the axis parameters $\chi_0$
  and $R_p$, and the time $t_0$ at which the shower passes the point of
  closest approach. (Image from D. Kuempel).}
\label{fig:SDP}
\end{centering}
\end{figure}

I have taken the PMT trigger times from Fig 3 of \cite{Hara}, and using an
estimate of the PMT pointing directions, I have made fits to the standard timing equation,
$$ t_i = t_0 +\frac{R_p}{c}\tan\left(\frac{\chi_0-\chi_i}{2}\right) $$ 
to extract shower axis parameters $t_0$, $R_p$ and $\chi_0$ from the
eight $(\chi_i,t_i)$ data points.  I have assumed a vertical
shower-detector plane (SDP), and I have guessed at a timing uncertainty of
0.05$\mu$s for each point.  The SDP and the axis parameters are illustrated
in Figure~\ref{fig:SDP}.

\begin{figure}[p]
\begin{centering}
\includegraphics[width=14cm]{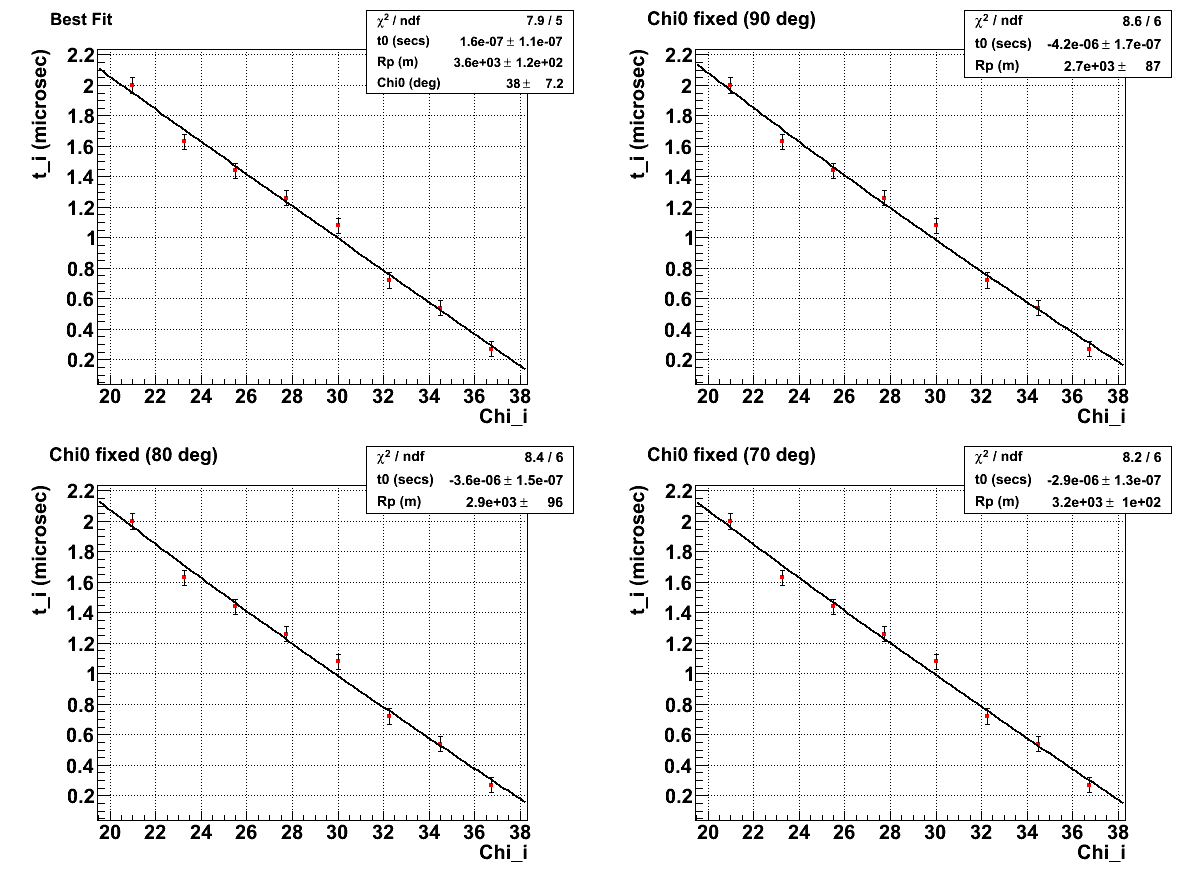}
\includegraphics[width=14cm]{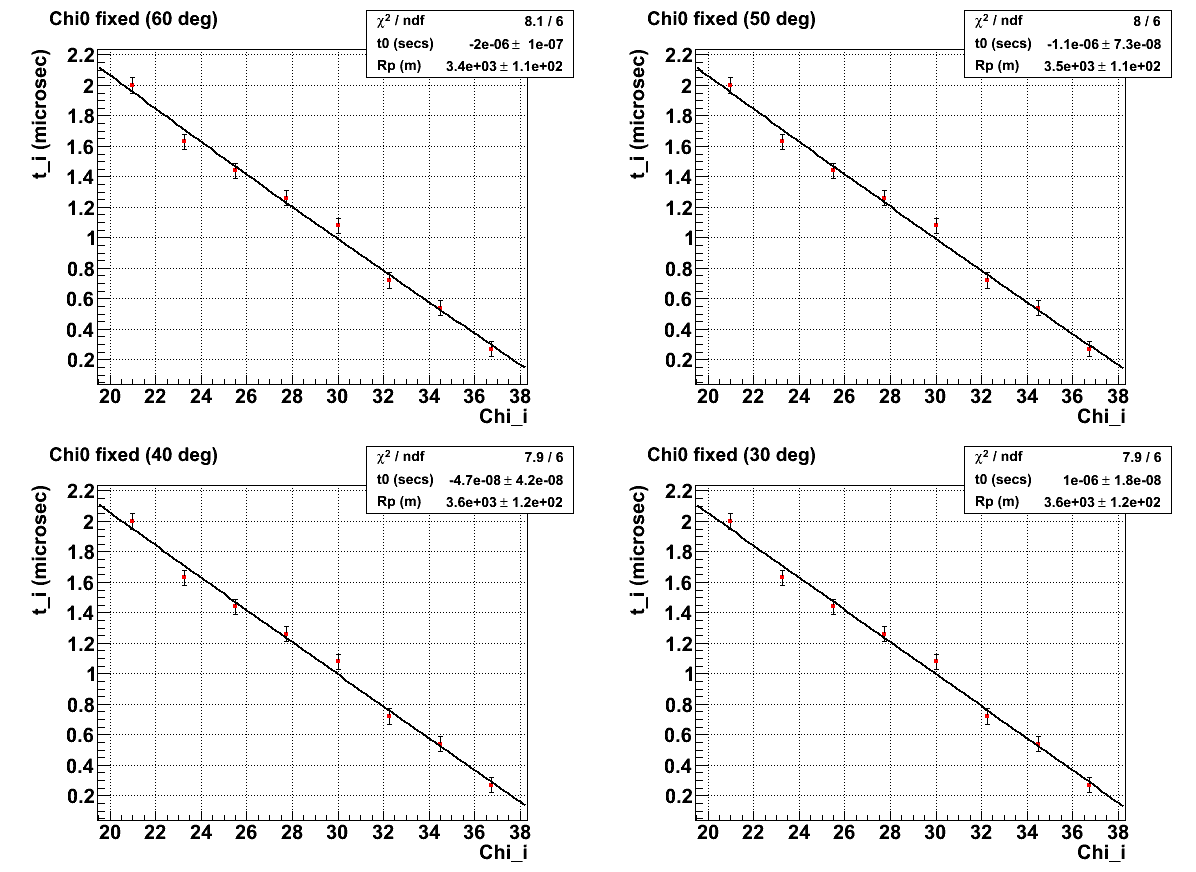}
\caption{Timing fits for the event.  We have assumed a vertical
  shower-detector plane and timing uncertainty of 0.05$\mu$s for each
  point.  The top left figure shows the best fit for $t_0$, $R_p$ and
  $\chi_0$.  The remaining plots show results when $\chi_0$ is fixed
  as indicated, and a fit is done for $t_0$ and $R_p$.  The short
  track length of the event precludes a unique reconstruction of the
  shower axis.}
\label{fig:time}
\end{centering}
\end{figure}

Results of various timing fits are shown in Figure~\ref{fig:time}.
Because of the rather short angular track length of the event
($18.4^\circ$), the timing fit suffers a large degeneracy in the
parameters $R_p$ and $\chi_0$; there is no curvature evident in the
$t_i$ versus $\chi_i$ plot, meaning that while we do have an estimate of the
the angular speed $\omega$ of the light spot across the camera, we have no
information about $d\omega/dt$.  The {\em best} fit in
Figure~\ref{fig:time} returns $\chi_0=38^\circ$ and $R_p=3.6$\,km, but
we show that a wide range of values of $\chi_0$ give acceptable fits.
Other values not shown (e.g. $\chi_0 > 90^\circ$) also give reasonable
fits.

The implication of this degeneracy is that the axis geometry of this
event is very uncertain - the shower could be vertical
($\chi_0=90^\circ$ assuming a vertical SDP) with an impact parameter
of $R_p=2.7$\,km, or the shower could be approaching the detector with
a zenith angle of $60^\circ$ (ie $\chi_0=30^\circ$) and $R_p=3.6$\,km.

The latter geometry of an inclined, approaching shower would produce a
light signal dominated by Cherenkov light.  If this were the case, the
claim of Hara et al. of the observation of fluorescence light would be
in question.  However, Hara et al. correctly point out that there is
important information in the {\em shape} of the light profile recorded
by the telescope.  We test this in the next section.

\section{Light Profile}

Figure~3 and Figure~4 of \cite{Hara} give information on the flux of
light received by the telescope as a function of time (or angle
$\chi_i$).  The key point is that the flux profile is rather flat.  I
have performed some simulations of a shower with a range of axis
geometries consistent with the timing fits from the previous section.
The simulated shower had a fixed energy of $5\times10^{18}$eV and a
depth of maximum $X_{\rm max} = 680$\,g/cm$^2$.  The aim of the
exercise is {\em not} to fit the observed light profile, but to
illustrate the change in the light profile shape as a function of the shower
geometry.

\begin{figure}[p]
\begin{centering}
\includegraphics[width=14cm]{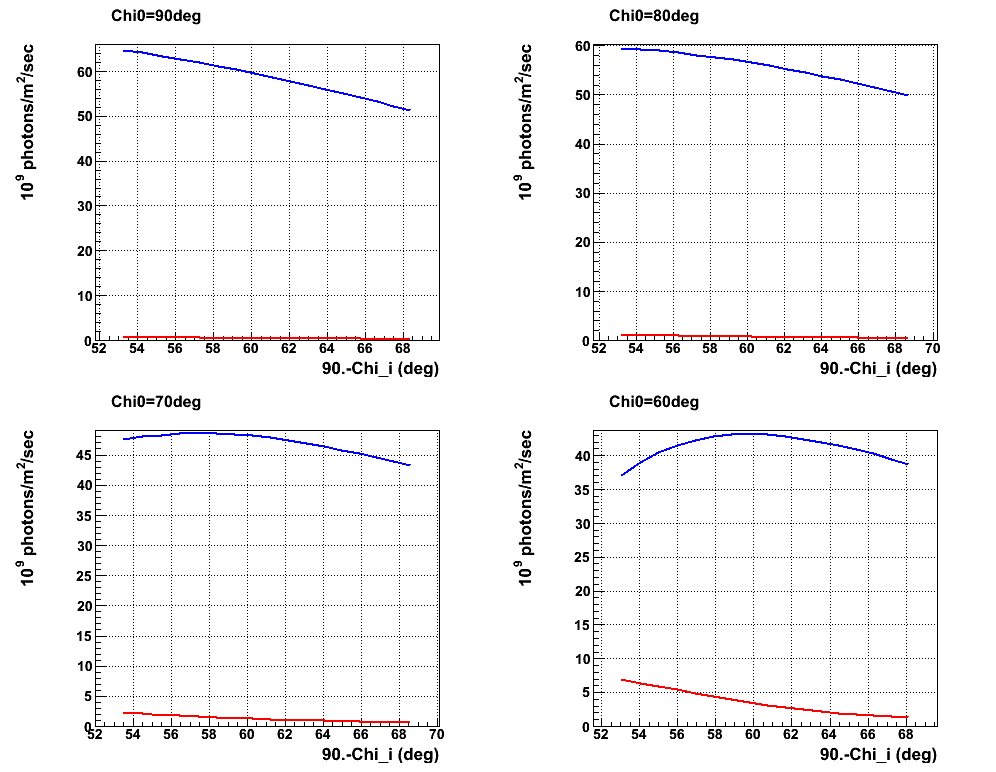}
\includegraphics[width=14cm]{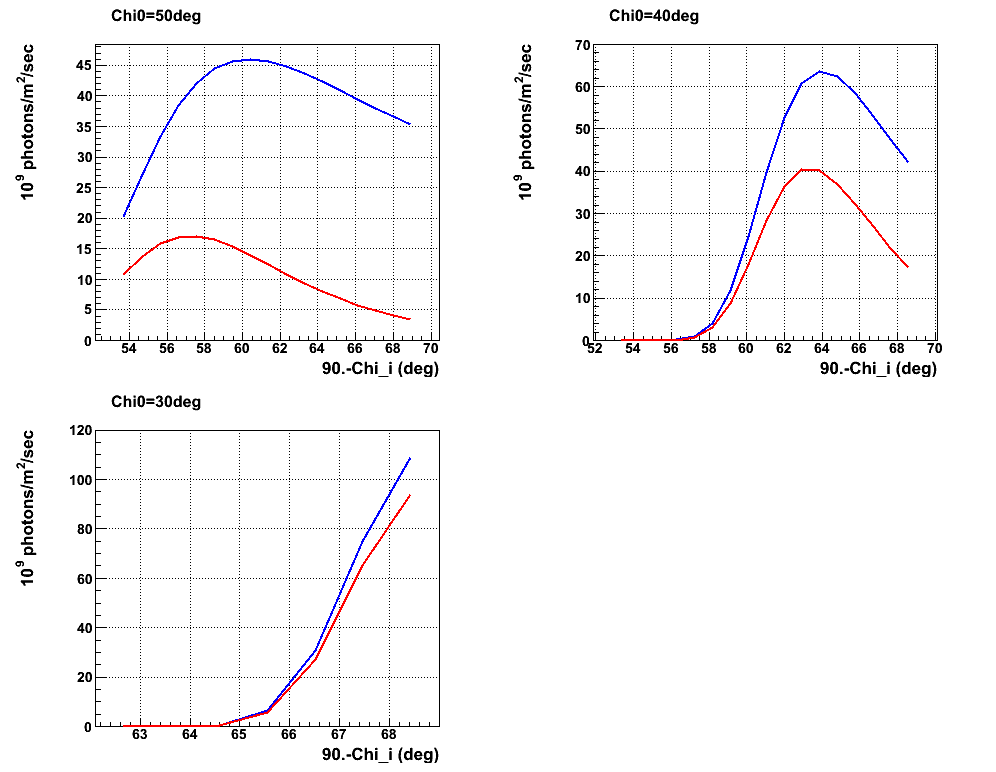}
\caption{Light flux at the telescope for a $5\times10^{18}$eV, $X_{\rm
    max} = 680$\,g/cm$^2$ shower for various axis geometries allowed
  by the timing fit. The x-axis represents the zenith angle of the
  light spot, given the assumption of a vertical SDP.  Blue lines
  indicate total light flux, and red lines show the contribution from
  direct Cherenkov light.  The shape and intensity of the light
  profile is a strong function of the event geometry.  The flatter
  light profiles in the first four panels are a better match to the
  observed light profile.}
\label{fig:flux}
\end{centering}
\end{figure}

Figure~\ref{fig:flux} shows the results of these simulations.  We find
that flat light profiles are only seen for more vertical showers.
Showers that approach the telescope ($\chi_0=50^\circ$ and smaller)
produce peaked light profiles, and produce signals contaminated by a
large direct Cherenkov light component.  The peaked shape of these
profiles is a consequence of the narrow angular distribution of
Cherenkov light around the shower axis.

\section{Conclusion}
It appears very likely that the signal detected by Hara et al. was one
dominated by fluorescence light, and it is reasonable that they lay
claim to the first such detection.  While the timing information of
the event is not sufficient to reconstruct a unique shower axis
geometry, the event's light profile is consistent with a fluorescence
profile, and quite unlike the peaked profile expected for a Cherenkov
dominated signal.

Of course some doubt does remain, and the first {\em unambiguous}
detection of fluorescence light from an air shower was with the Utah
Fly's Eye prototype operated in coincidence with the Volcano Ranch
surface array in 1977 \cite{FE}.

The $5\times10^{18}$eV shower simulated in this note produced a light
profile with a peak flux of about $5\times10^{10}$\,photons/m$^2$/s,
the flux observed in \cite{Hara}.  This energy is a factor of two
lower than that claimed by Hara et al., but this can be explained by
their assumption of a fluorescence yield of 1.9 photons per metre of
electron track, which is approximately a factor of two lower than the
modern accepted value.

One remaining uncertainty in this energy is the absolute light
calibration of the experiment.  Hara et al. discuss an inconsistency
between the observed light intensity spectrum and that expected from
simulations.  The inconsistency led to them increasing the observed
light fluxes by a factor of 5.  The origin of the inconsistency is
unclear, but it would probably be safe to say that $5\times10^{18}$eV
is an upper limit to the energy of the observed air shower.

\section{Acknowledgements}
I'm grateful to Alan Watson for encouraging me to look into these experimental
results, and to Greg Thornton for some helpful discussions.

\end{document}